\documentclass{article}

\usepackage{PRIMEarxiv}

\usepackage[utf8]{inputenc} 
\usepackage[T1]{fontenc}    
\usepackage{hyperref}       
\usepackage{url}            
\usepackage{booktabs}       
\usepackage{amsfonts}       
\usepackage{nicefrac}       
\usepackage{microtype}      
\usepackage{lipsum}
\usepackage{fancyhdr}       
\usepackage{graphicx}       
\graphicspath{{media/}}     
\usepackage{listings}       
\usepackage{changepage}
\title{OpenAPI Specification Extended Security Scheme: A method to reduce the prevalence of Broken Object Level Authorization
}

\author{
  Rami Haddad \\
  Cisco OutShift \\
  Amsterdam\\
  \texttt{r-h\_rami@live.nl} \\
   \And
  Rim El Malki \\
  Cisco OutShift \\
  École Polytechnique \\
  Paris\\
  \texttt{reem.melki.207@gmail.com} \\
    \And
  Daniel Cozma \\
  Cisco OutShift \\
  Bucharest\\
  \texttt{scozma@cisco.com} \\
}

\begin{document}
\maketitle

\begin{abstract}

APIs have become the prominent technology of choice for achieving inter-service communications\cite{PostmanStats}. The growth of API deployments has driven the urgency in addressing its lack of security standards. API Security is a topic for concern given the absence of standardized authorization in the OpenAPI standard, improper authorization opens the possibility for known and unknown vulnerabilities, which in the past years have been exploited by malicious actors resulting in data loss. This paper examines the number one vulnerability in API Security: Broken Object Level Authorization(BOLA), and proposes methods and tools to reduce the prevalence of this vulnerability. BOLA affects various API frameworks, our scope is fixated on the OpenAPI Specification(OAS). The OAS is a standard for describing and implementing APIs; popular OAS Implementations are FastAPI, Connexion (Flask), and many more \footnote{\url{https://github.com/OAI/OpenAPI-Specification/blob/main/IMPLEMENTATIONS.md}}. These implementations carry the pros and cons that are associated with the OAS's knowledge of API properties. The Open API Specification’s security properties do not address object authorization and provide no standardized approach to define such object properties. This leaves object-level security at the mercy of developers, which presents an increased risk of unintentionally creating attack vectors. Our aim is to tackle this void by introducing 1) the OAS ESS (OpenAPI Specification Extended Security Scheme) which includes declarative security controls for objects in OAS (design-based approach), and 2) an authorization module that can be imported to API services (Flask/FastAPI) to enforce authorization checks at the object level (development-based approach). When building an API service, a developer can start with the API design (specification) or its code. In both cases, a set of mechanisms are introduced to help developers mitigate and reduce the prevalence of BOLA. \\

\end{abstract}

\pagebreak
\section{Introduction}
\noindent One key requirement for achieving strict security in any system is a deny-by-default access property where privileges are granted on a need-to basis per request, adhering to a security principle commonly known as the Principle of Least Privilege (PoLP)~\cite{jero2021practical}. PoLP is a vetted industry standard today and remains the preferred access control approach; commercially referred to as the zero-trust model~\cite{ahmed2020protection}. APIs present a wide array of entry points for malicious actors; given the absence of standardized API security controls regarding users and objects metadata, it can be concluded that APIs do not adhere to the PoLP leaving the data managed by the API service "at-risk". Whilst many custom frameworks and $3^{rd}$ party services exist to address object access control, there is no standardized native security property that handles declarative security coding of authentication and authorization functionalities for API objects. The OpenAPI Specification (OAS)~\cite{schwichtenberg2017open} outlines a standard for API design, including certain user authentication security properties. The OAS security scheme does not address (object-level) access control for users or groups, resulting in the absence of defined object-level authorization rules which leaves object metadata in a vulnerable state with a possible breach of data confidentiality. Based on this design, it can be concluded that the OAS does not adhere to the PoLP. Broken Object Level Authorization (BOLA)~\cite{viriya2021peeking} also known as Insecure Direct Object References (IDOR) is a key security issue listed in the OWASP API\footnote{\url{https://owasp.org/www-project-api-security/}} top 10 security/vulnerability issues list for API services. BOLA cases are cases where unauthorized users can access restricted objects implemented by an API service. BOLA is a symptom of sloppy programming: BOLA cases often occur when the programmer incorrectly programmed the object authorization code (if at all) for objects that are addressed. A common flaw in API design is the use of numbers as resource IDs; non-unique elements allow for guessing and present an initial flaw in object vulnerability. BOLA typically represents a large attack surface and proves to be difficult in detecting and mitigating given the lack of contextual knowledge of objects. The OAS does not describe objects rendering the API service unaware of objects' existence and their associated interactions with users, this presents an obstacle in the creation of an object-based permissions model. Achieving object-level authorization is also hampered by the OAS inability of object awareness at the design level, this trait is shared among all Interface Description Languages (IDLs)~\cite{snodgrass1989interface} and presents a flaw in terms of enforcing rules on API metadata; objects for instance. Object permissions require a system-level awareness of objects and the existence of a security mechanism for object-level permissioning. Interface Description Languages (IDL) such as the OAS are unaware of objects leaving the developer with no properly defined method to describe its attributes, including authorization. IDL’s current state presents an issue in the design process of API security. On the other hand, the implementation and run-time environments are object-aware.\\
 
\noindent Alternatively, a remodeled OAS based on OWASP C7\footnote{\url{https://owasp.org/www-project-proactive-controls/v3/en/c7-enforce-access-controls}} proactive controls principles would favor clear object definitions and an authorization mechanism enforcing object access control. Following C7 proactive controls principles in API design will be impactful on the “development stage” where most-often developer deficiency in security-knowledge results in the unintentional creation of attack vectors. A suggested “fix” for the broken authorization in object permissioning is the extension of the OAS Security Scheme (OAS ESS) to standardize a mechanism that allows for object-based authorization in run-time through declarative security descriptions. The argument for this fix is based on the removal of authorization complexity from the developer to the “background”, exposing only declarative security controls to the developers. When building an API service, a choice is presented to the developer to start with the API design by defining an API definition, or to utilize OAS implementations heading straight to code development. In an API design-first approach, the effects of the proposed Extended Security Scheme will guide the developer in defining object authorization, effectively “auto-creating” the ACLs and their associated objects/groups in a generated server stub. Assuming a developer approach, where the API definition file (specification) is generated after the API service is coded, the OAS ESS documentation serves as a set of guiding principles for developers providing the capability of declarative security controls for objects, with its complexity hidden in the root files of the programming environment.\\

\noindent In summary: (1) OAS needs to provide and standardize an authorization capability to secure resources from unauthorized access, (2) however the OAS needs additional methods to attain object knowledge and provide multi-user object authorization privileges and this paper addresses the gamut from 1 to 2.\\

\pagebreak

\section{OpenAPI Specification Extended Security Scheme}
\label{sec:headings}
The Open API Specification does not define objects, leaving API services with no ability to define object properties e.g., Access Control Lists. This leads to BOLA issues. In this CPOL, we present a set of methods addressing common developer flaws that can reduce/prevent the unintentional creation of BOLA attack vectors in API services. We approach this issue at both the design level and during code construction. In the first approach (design-level), we extend the IDL to enforce object-level authorization (OAS ESS). In the second approach (code construction), we introduce an authentication and authorization module that can automatically (or semi-automatically in some cases) generate ACLs for object resources based on user groups (RBAC) and user IDs. The presented module is a package that can be imported to enable authorization in services/applications that lack it. Finally, we introduce a generator that can detect the existence of the proposed declarative-security mechanism. This generator works both ways; it can generate a specification from a given API service or vice versa.

Interface Definition Languages (IDLs) are commonplace in language-independent systems and are predominantly used to facilitate remote procedure calls (RPCs), one of which is API calls. The metadata of an API service lives in run-time and does not reflect its state of resources to the IDL; (OAS). This renders the specification unaware of the consumed objects and their details which makes it difficult to describe any properties to metadata, including objects in design-time. As a result, object authorization cannot be defined in the code leaving the metadata of an API service at risk of unauthorized operations. BOLA is an API security risk that occurs when legitimate (authenticated) users succeed in accessing or modifying data in an unauthorized state. One method to address BOLA would be the creation of ACLs that describe a set of rules on object consumption. However, these are difficult to implement and leave room for sloppy programming, which is a major cause of BOLA. BOLA typically appears when developers write and use sloppy/recycled code whilst lacking thorough code security knowledge to ‘close the gaps’ and prevent BOLA scenarios.\\

\noindent BOLA presents itself as much as a human problem, as it is a technical problem. It is not feasible to achieve the best-practice scenario in which every developer is thoroughly aware of BOLA and ensures that the code meets the requirements to not present BOLA. However, it certainly does minimize the risk significantly when developers can define object security and review the paradigm. Enabling developers in doing so, the complexity of authorization is to be taken to the background of the respective programming environment, whilst presenting a declarative-code entry for developers which automatically generates the associated ACLs for object access. Such an approach includes enforcing properties on the metadata of an API service; its objects, which leads to the requirement of solving or finding a workaround to the limitation of IDLs.\\

\noindent To overcome these issues, we have developed a simple mechanism to 1) enforce object authorization checks in API services. We have also proposed 2) a security extension scheme for OAS and 3) a tool that recognizes these extensions, going from specification to code and code to specification.\\

\begin{figure}[h!]
  \begin{center}
  
  \includegraphics[scale=0.25]{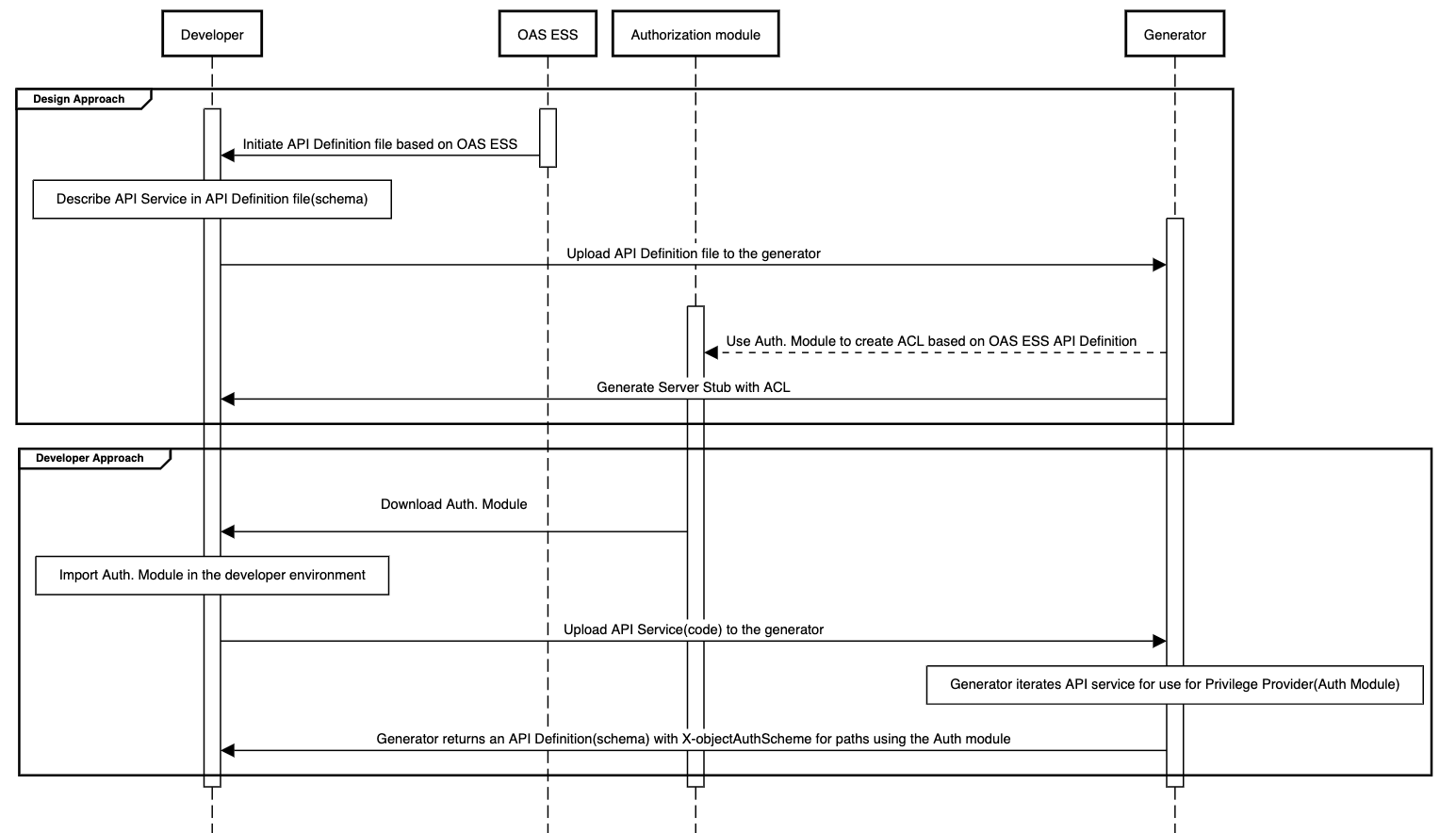}    \caption{Extended Security Scheme call flow}
    \label{Call Flow}
  \end{center}
\end{figure}
\newpage

\subsection{Evolving the Specification}


\noindent Consider two designs for object-level authorization checks in the OAS. The first design which defines object security, globally (root-level),  in the ‘securitySchemes’  is shown below:\\

\begin{adjustwidth*}{0cm}{-0.4cm}
\begin{lstlisting}[caption=Root-level security | OAS ESS]
-- components:
  schemas: {...}
  securitySchemes:
    api_key:
	    in: header
	    name: api_key
	    type: apiKey
    X-objectAuthScheme:
	    type: apiKey
	    name: api_key
	    in: header
	    x-groups: string
	    x-user_id: string	
\end{lstlisting}
\end{adjustwidth*}

\noindent The X-objectAuthScheme is now defined using the type apiKey for demonstration purposes (can be JSON web tokens). The X-objectAuthScheme can be implemented at the method level (defined for every method in every path) or at the root level (globally defined in ‘securitySchemes’ and referenced in every path). This design represents the latter case, and it utilizes the components -> securitySchemes structure which keeps security definitions at the root level of the API Definition. The ‘components’ section in OAS includes common definitions that are used by multiple API operations. To avoid duplication, these definitions are defined globally in ‘components' and are referenced using ‘\$ref’ (reusable definitions). An OAS generator will require adjustments to recognize this proposed security scheme. Next, we include ‘X-objectAuth’ for each object in components --> schemas and ‘X-objects’ for each path to enforce object-level authorization.\\
\pagebreak

\begin{adjustwidth*}{0cm}{-0.4cm}
\begin{lstlisting}[caption=Applying Root-level Security | OAS ESS]
--paths:
	/pet:
	  post: 
	    requestBody: {...}
	    responses: {...}
	  put:
	    requestBody: {...}
	    responses: {...}
	  x-objects:
	    $ref: '#/components/schemas/Pet/x-objectAuth'

-- components:
	schemas:
	  pet:
	    type:
	      object
	    properties:
	      id:
	        type: integer
	        format: int64
	      name:
	        type: string
	        example: lucky
	    x-objectAuth:
	      object:
	        $ref: '#/components/schemas/Pet/properties/id'
	      schema:
	        $ref: '#/components/securitySchemes/X-objectAuthScheme'
	      scopes:
	        groups:
	          $ref: '#/components/securitySchemes/X-objectAuthScheme/x-groups'
	        user_id:
	          $ref: '#/components/securitySchemes/X-objectAuthScheme/x-user_id'
	        methods:
	          post:
	            description: create an object
	          put:
	            description: modify/update an object
	          get:
	            description: read an object
	          delete:
	            description: delete an object
\end{lstlisting}
\end{adjustwidth*}

\noindent The second design which is at the method-level is shown below: \\

\begin{adjustwidth*}{0cm}{-0.4cm}
\begin{lstlisting}[caption= Applying Method-level Security | OAS ESS]
--paths:
	/pet:
	  post:
	    requestBody: {...}
	    responses: {...}
	    X-objectAuth:
	      object: 
	        schema: 
	          $ref: 'post/responses/'201'/content/application/json/schema/properties/identifier'
	      token:
	        type: JWT
	        name: JSON web token
	        in: header
	      scopes:
	        C:
	          groups:
	            type: string
	          user_id:
	            type: string
	        R:
	          groups:
	            type: string
	          user_id:
	            type: string
	        U:
	          groups:
	            type: string
	          user_id:
	            type: string
	        D:
	          groups:
	            type: string
	          user_id:
	            type: string
\end{lstlisting}
\end{adjustwidth*}

\noindent The downside of this design is that there will be redundant information across the specification since ‘X-objectAuth’ will be multi-listed within every method in each path (duplicate information). Choosing a root-level design or method-level design depends on the developer and the requirements of the API service.\\

\subsection{Automatic creation of ACLs by the authentication and authorization module}
The goal of the authorization module is to enforce authorization at the object level in API services. It is an “add-on” feature that can be imported into any API service/application. This module automatically generates ACLs based on predefined rules. Rules are specific to each service/application, and they are based on user roles.  

\begin{figure*}[ht!]
\centering
\includegraphics[width=0.65\textwidth]{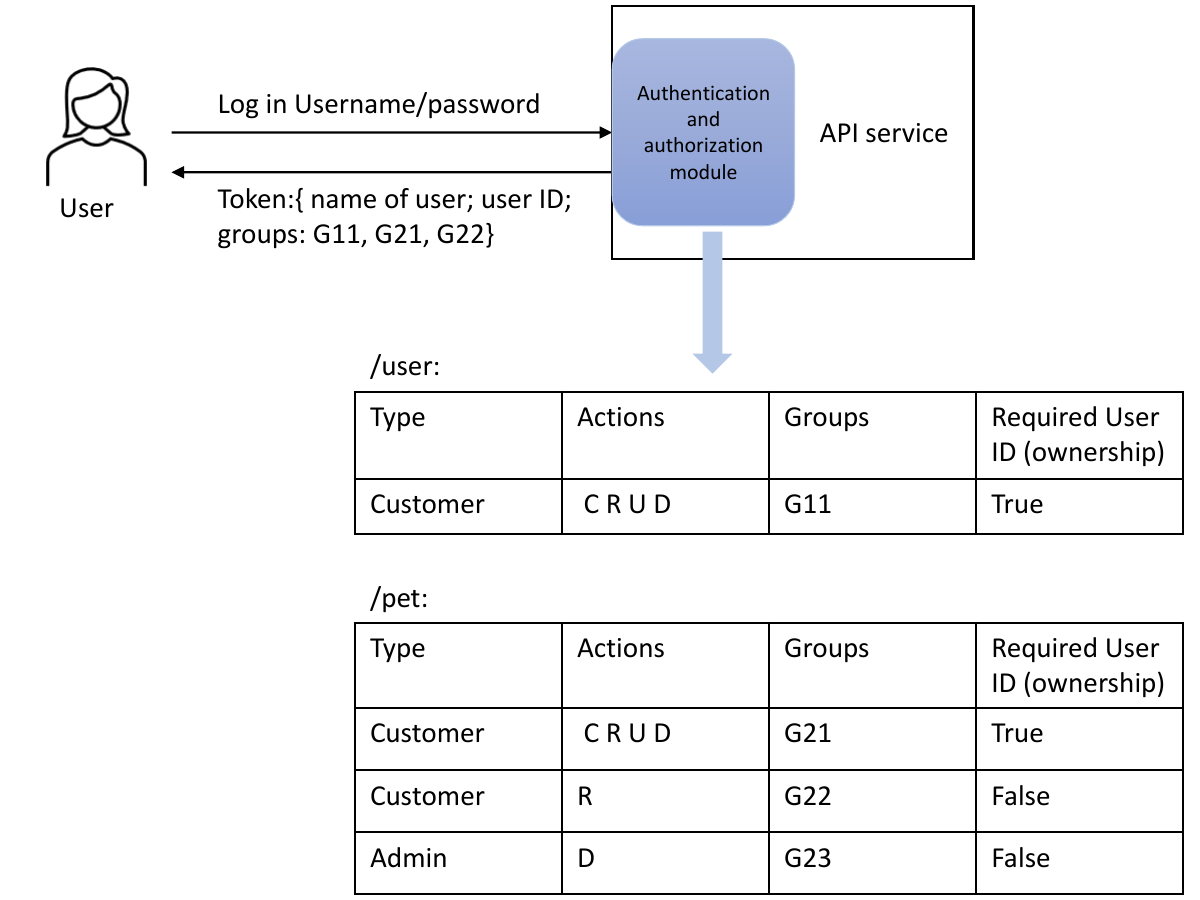} 
\caption{Authentication mechanism}
\label{authmec}
\end{figure*}

Following successful authentication, a user is granted a token that includes the user’s name, unique ID, and group(s) (Fig.~\ref{authmec}). Groups are based on the type of user (e.g., customer, admin) and they are path-based. This means that for every path in the API service, groups and their associated actions are listed. An example is shown in Fig.~\ref{authmec}. The API service in this example has two paths (‘/user’ and ‘/pet’). In the ‘/user’ path, users belonging to group ‘G11’ can create, read, update, and delete their own objects (since ownership is true/required). In the ‘/pet’ path/directory, users belonging to group ‘G21’ are also able to create, read, update, and delete their own objects. Users belonging to group ‘G22’ can read any object in the ‘/pet’ directory since object ownership is not required. Users belonging to group ‘G23’ can delete any object in the ‘/pet’ directory since object ownership is not required. For each API service, a different set of rules are tailored to match the functionality of the service and the type of users utilizing this service. Following successful authentication, a user can start interacting with the API service using HTTP queries.  

\begin{figure*}[ht!]
\centering
\includegraphics[width=0.85\textwidth]{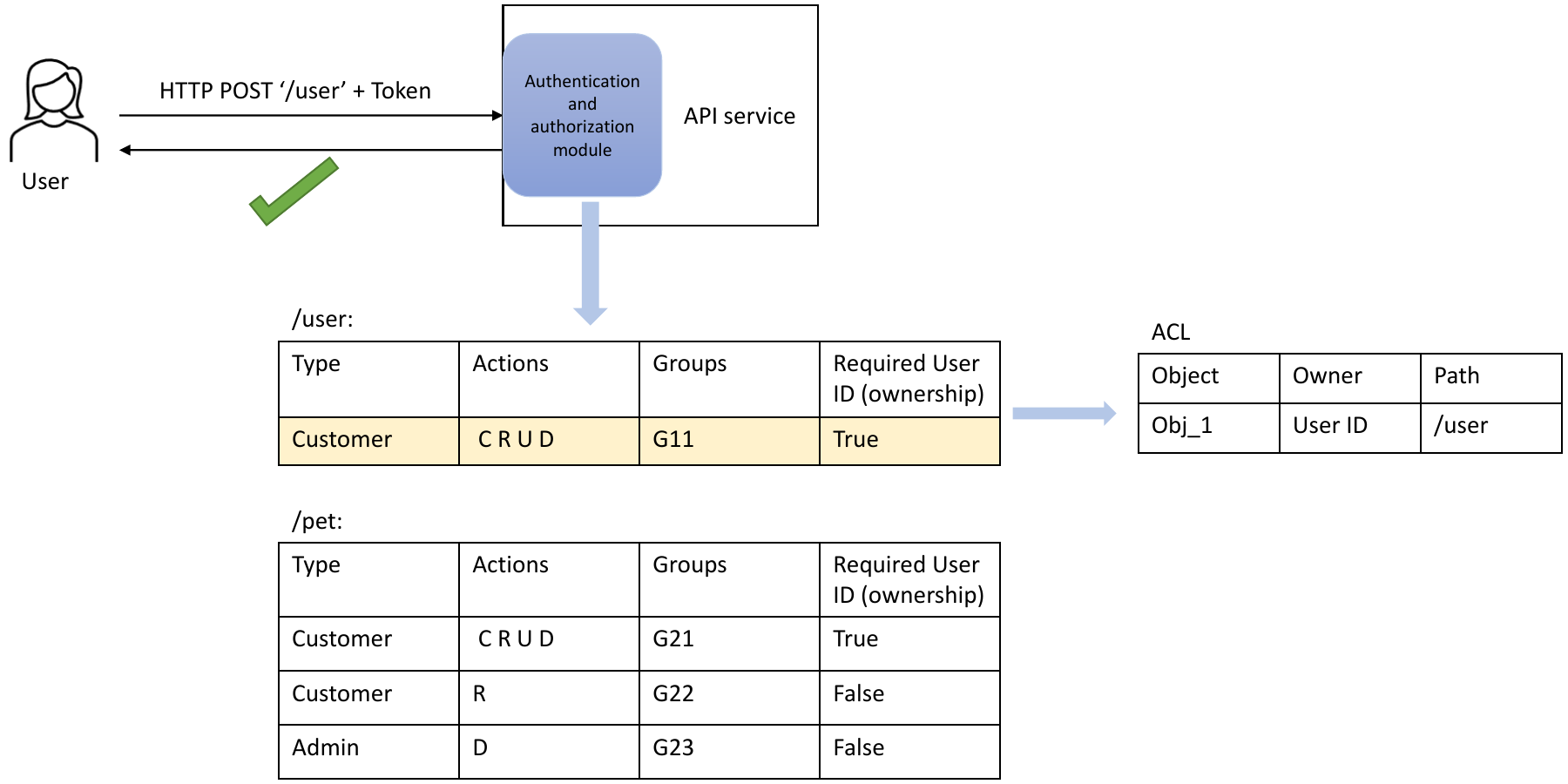} 
\caption{An authorized object creation mechanism}
\label{authz}
\end{figure*}

Figure~\ref{authz} shows an example of a create request (POST). The user sends a POST request containing its token to the API service in order to create an object in the path ‘/user’. The authentication and authorization module first examines the group ID included in the token and verifies whether the user is able to create an object (POST) in the requested path. If this information is valid, the module extracts the user ID (because it is required in a create process) from the token and creates an entry in the ACL that includes the object ID and the user ID. Assuming that the user wants to modify the created object by sending a PUT request. Again, the module examines whether the user belongs to a group that is authorized to update objects in the identified path. If this information is valid, then the module examines the ownership requirement for that particular group. If ownership is required, then the module extracts the user ID from the sent token and compares it to the user ID that is associated with the object ID in the ACL. If it matches, then the user would be able to update the object. It should be noted that if ownership is false or not required then a user would be able to update, read and delete objects without being the owner of these objects. If a user belongs to two different groups (e.g., G21 and G22 in the path ‘/pet’) that include the same action (R) but have different ownership requirements (one is true and the other is false), then false ownership overrides the true one. This is attributed to the fact that a user will be to read, update, and /or delete any object in a specific directory including his own. Using this approach, ACLs will be automatically generated (users will be associated/linked to objects). The generated ACLs can be saved in permanent storage since APIs are stateless.  

\subsection{Generator / Privilege Provider}
\noindent To overcome the limitation of IDLs, an ‘object-aware’ mechanism is introduced utilizing object identifiers. This mechanism is implemented using a newly (OAS ESS) introduced security Scheme; ‘X-objectAuth’. Using this keyword in the specification, or its corresponding privilege provider function in the API service has a direct effect on the generator and its output.\\

\noindent In this implementation, a custom-built generator is used to prove the ability to enforce code-level object authorization through declarative programming. This generator can create a server stub from a specification and vice-versa.\\

\textbf{Specification to API Server Stub }

In a scenario where the generator takes an API Specification to generate an API server stub, the API specification is parsed by the generator, retrieving the existing paths and their HTTP verbs. The presence of the given keyword (X-objectAuth) is checked per path which indicates the decision to enforce object authorization. Subsequently, the generator checks if ‘X-objectAuthScheme’ exists within the listed securitySchemes. In case of such presence, the generator adds the privilege provider module to the server stub. The generator checks the details of ‘X-objectAuthScheme’ in securitySchemes (‘in: header’, ‘x-groups’, ‘x-user\_id’, ….) and matches it to the available module (parsing the module code).\\

\textbf{API Server Stub to Specification}

In a scenario where the generator takes an API server stub to generate the specification, the server stub code is parsed by the generator and its lines are iterated for the presence of the ‘privilege provider generates object privilege’ wrapper function which is defined as a decorator for each existing path. In the case of such presence, the generator creates the ‘X-objectAuth’ property in the specification. The privilege provider is read to obtain the existence of Group ID (x-groups), User IDs (x-user\_id) checks in the module, and where its associated JWT is located (body/header).\\

Read and Read/Write privileges are stored in an ACL format (the previously presented ACL design can be extended to include user IDs). The ACL in the server stub contains Access Control Entries (ACE) which define the users' read(R) and write(W) privileges. Given is an example ACE, where an object with ID {152} extends RW privileges to user ID {123}. User ID 123 happens to be the owner of this object, as every object is automatically assigned an owner (the creator of the object). 
\pagebreak

\begin{adjustwidth*}{0cm}{-0.4cm}
\begin{lstlisting}
  { 
    "id": 152,     
    "path": "/pets", 
    "owner": "123", 
    "users_ro": [], 
    "users_rw": [ 
      "123" 
    ] 
  }	
\end{lstlisting}
\end{adjustwidth*}

\section{Conclusion}
As counter mechanisms evolve, so do their adversaries. This is the unchanging truth that keeps generations in a problem-solving infinite loop, or is it not? In theory, future endeavors may undertake techniques that can truly generate hack-proof systems, as programming paradigms evolve and battle for the preferences of their audiences. The quest to craft a programming system, designed to be resistant to hacks results in arguments that date back to the days of Auguste Kerckhoffs, who advocated against the principle of Security through Obscurity. As open source solutions have gained traction since the adoption of Unix, it is no small feat to convince a community into Security through Obscurity as a viable approach, neither is this an advocated principle in this paper. It is more effective to disrupt habits detrimental to Kerckhoff's principles of the properties that make a system, secure. Securely-architectured systems are uncommon, despite the 'Secure by Design' claims of numerous applications constructed with programming languages designed with security as an afterthought. Computing in its very foundation functions on I/O(Input/Output) signals, this is also where security tends to fail by design. Securely-architectured systems carry the property that if and only if the input to a system results in a defined(logical) and desired behavior, it can be mathematically secure. This is also known as formal verification; proving or disproving the truth of intended algorithms. It may be argued that mathematically verified systems will end the infinite loop of cybersecurity conflicts, whilst unlikely, said securely-architectured systems are likely to have significant defying effects on developers' sloppy programming.\\

\noindent Till such inventions make it to daylight, a 'stop-gap' fix is required for immediate use that can alleviate a developer's inability in spotting code vulnerability during API construction, and API design, to ultimately reduce the prevalence of Broken Object Level Authorization. BOLA is not a result of a weak architectured system, but rather weak implementations by developers. Poor Code Construction is a widespread problem and has been discussed by Dijkstra, as well as McConnel whose work has had a large impact on software design\cite{ReliabilityPrograms,codeComplete}. By introducing the OAS ESS, the developer is given a declarative-programming front to define and apply authorization rules in the form of an ACL for objects. This has two direct consequences; the chance for sloppy programming is significantly reduced as authorization logic is coded in a module that has been peer-reviewed by the open-source community, and developers are 'guided' through an authorization declaration process which yields educational results, as developers gain insight on properly defining object access rules.\\

\noindent The OAS ESS remains to be put to the test in API deployments. Its scalability is untested, and there is no data to determine how the ESS will behave in the scenario of a malicious user intending to misuse the Authorization Provider or manipulate the ACL. In theory, this mechanism may introduce a new array of vulnerabilities. Taking tasks/complexity away from the developers is not a new thing, most programming endeavors today are fast as libraries exist for nearly any (complex) task. The OAS ESS may require additional remodeling, a solid implementation, and rigorous white and black box testing to yield the data that can answer our questions.\\

\noindent In this paper, the concept of declarative security controls for the OAS has been researched, which resulted in the OAS ESS. The OAS ESS has been presented at the ASC API Specifications Conference, in September of 2022 at San Francisco\footnote{\url{https://www.youtube.com/watch?v=FncoYX_jrwo}}. The aim is, to drive awareness surrounding (the lack of) object authorization, and introduce the concept of the OAS ESS, with the long-term aim of standardization. Further studies and implementations are encouraged to bring the OAS ESS concept to maturity.\\

\section*{Acknowledgments}
This research was sponsored by Cisco Systems, and guided by Peter Bosch - Distinguished Engineer at Cisco OutShift. 

\bibliographystyle{unsrt}  
\bibliography{references}

\end{document}